\documentclass[aps,prl,reprint,groupedaddress,showpacs]{revtex4-1}

\usepackage{graphicx}
\usepackage{dcolumn}
\usepackage{bm}
\usepackage{textcomp}

\usepackage{amsmath}   
\usepackage{paralist}  
\usepackage{physics} 

\usepackage{amssymb}

\newcommand{\overbar}[1]{\mkern 1.5mu\overline{\mkern-1.5mu#1\mkern-1.5mu}\mkern 1.5mu}

\begin{document}

\preprint{APS/123-QED}

\title{Role of Optical Density of States in Two-mode Optomechanical Cooling}

\author{Seunghwi Kim}
\author{Gaurav Bahl}%
\email{bahl@illinois.edu}
\affiliation{%
	Department of Mechanical Science and Engineering, University of Illinois at Urbana-Champaign, Urbana, IL 61801, USA}%

\date{\today}

\begin{abstract}

Dynamical back-action cooling of phonons in optomechanical systems having one optical mode is well studied. Systems with two optical modes have the potential to reach significantly higher cooling rate through resonant enhancement of both pump and scattered light. Here we experimentally investigate the role of dual optical densities of states on optomechanical cooling, and the deviation from theory caused by thermal locking to the pump laser. Using this, we demonstrate a room temperature system operating very close to the strong coupling regime, where saturation of cooling is anticipated.

\end{abstract}


\maketitle

The cooling of nanomechanical resonators is important for exploring the quantum behavior of mesoscale systems, high precision metrology \cite{Teufel2009, KrauseWingerBlasiusEtAl2012, GavartinE.VerlotP.KippenbergT.2012}, and quantum computing~\cite{O’ConnellHofheinzAnsmannEtAl2010, BagheriPootLiEtAl2011}. 
Developments in laser induced passive optical feedback cooling~\cite{MetzgerKarrai2004} and dynamical back-action cooling via ponderomotive forces~\cite{{ArcizetCohadonBriantEtAl2006},{GiganBohmPaternostroEtAl2006},{Schliesser2006}} have enabled access to the mechanical ground state~\cite{SchliesserRiviereAnetsbergerEtAl2008, Park2009, Chan2011} and the exploration of quantum mechanical effects \cite{Leggett2002}. The cooling rate achievable by such optomechanical mechanisms depends on the optical density of states (DoS) available for light-scattering and also increases with the intensity of the cooling laser~\cite{Genes2008}.
Resonant enhancement of the pump, for instance through two-mode optomechanics, can thus increase the cooling rate by enhancing the number of intracavity photons. 
Such two-mode systems have enabled experimental demonstration of phonon lasing~\cite{Grudinin2010, Bahl2011a}, cooling~\cite{Bahl2011c}, induced transparency~\cite{Kim2015a} and are even theoretically predicted to allow access to the strong coupling regime~\cite{Ludwig2012a, Stannigel2012}.

\begin{figure}[t]
	\centering
	\includegraphics[width=0.48\textwidth]{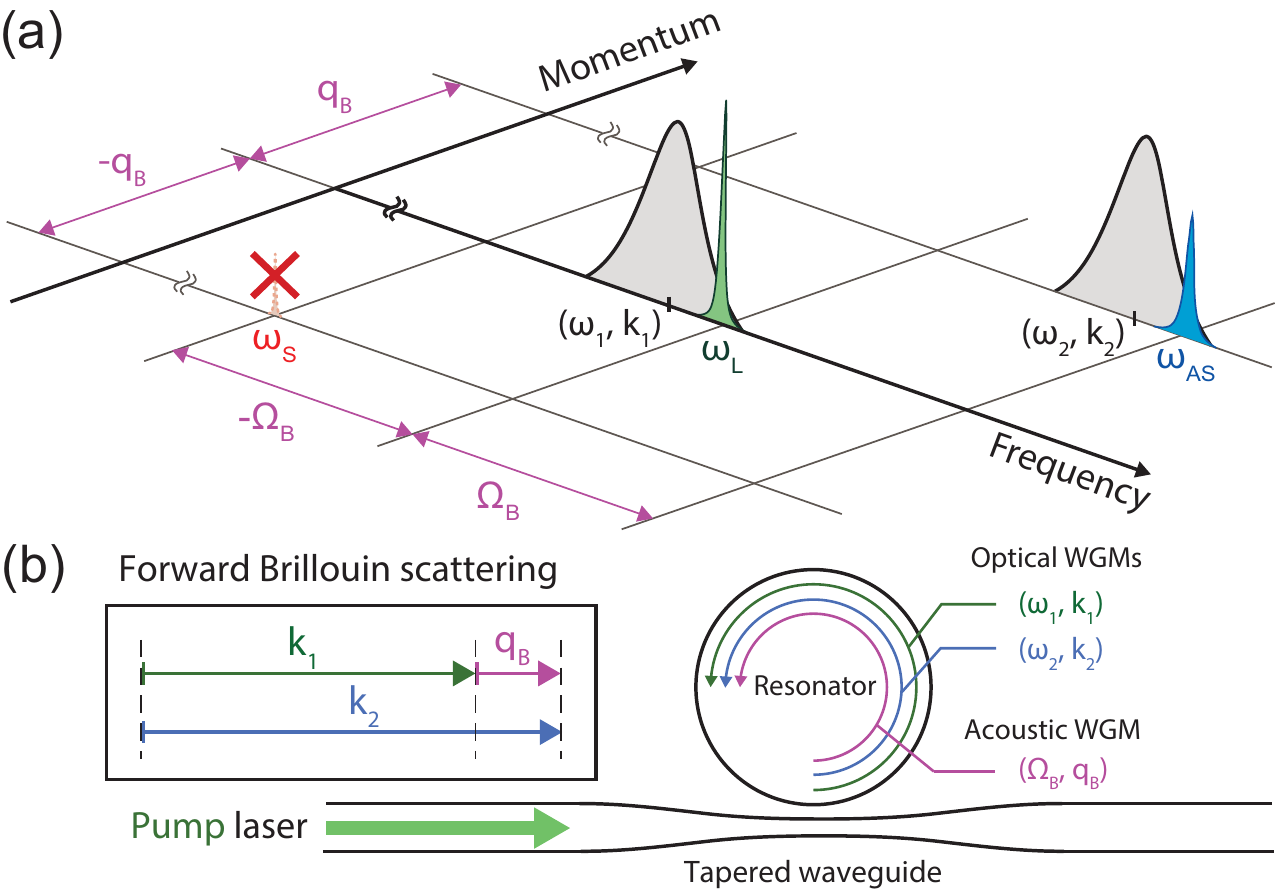}
	\caption{ (a) Generalized modal arrangement (phase matching diagram) required for dynamical back-action cooling of a ($\Omega_B, q_B$) phonon mode in a resonator supporting two optical modes ($\omega_1, k_1$) and ($\omega_2, k_2$). Phase matching necessitates consideration of both frequency and momentum separation. Here $\omega_{L}$ and $\omega_{AS} (\omega_{S})$ denote the frequencies of pump laser and the anti-Stokes (Stokes) scattered light. Stokes scattered light does not have a supporting optical mode and is thus suppressed.
		(b) Our experiments are performed in a whispering gallery microresonator that is optically coupled using a tapered fiber waveguide. Optomechanical cooling of an acoustic whispering gallery mode (WGM) having nonzero momentum $q_B \neq 0$ is achieved by means of forward Brillouin scattering. }
	\label{fig:1}
\end{figure}

Let us consider a general optomechanical system having two optical modes; the lower mode $\omega_1$ hosts the resonantly enhanced pump laser, while the higher mode $\omega_2$ hosts anti-Stokes scattered light. These two optical modes are coupled by a phonon mode of frequency $\Omega_B$ with the relation $\Omega_B \approx \omega_2 - \omega_1$. There are now two possibilities for the momentum associated with this phonon mode. In conventional optomechanical cooling the resonator supports standing-wave phonons, such as in a breathing mode, having zero momentum $q_B = 0$. Alternatively, a whispering gallery resonator may support a traveling wave phonon mode with non-zero momentum $q_B \neq 0$. For phase matching either process, momentum conservation $k_1 + q_B = k_2$ is necessary, which we illustrate in general in Fig.~\ref{fig:1}.
When the phase matching condition is satisfied, resonant pump light can scatter into the anti-Stokes optical mode ($\omega_2, k_2$) while annihilating phonons from the system. Stokes scattering is simultaneously suppressed since no suitable optical mode is available. This leads to the insight that the resolved sideband regime could also be achieved in momentum space, even if the optical modes are not resolved in frequency space, enabling cooling with high fidelity selection of only anti-Stokes scattering.

The existence of two-mode systems with $q_B \neq 0$ phonon modes was experimentally verified recently \cite{Bahl2011a}. Optomechanical cooling in these systems has been experimentally demonstrated \cite{Bahl2011c} and also theoretically analyzed \cite{Tomes2011, Agarwal2013}.
However, the expected role of optical DoS in the cooling process has not been experimentally investigated. 
In this work, we study the dependence of cooling rate on both pump and anti-Stokes optical DoS using the Brillouin opto-acoustic cooling system \cite{Bahl2011c} using a silica whispering gallery resonator. 
We analyze the role of thermal locking \cite{Carmon2004} and also discuss through experiment the potential of reaching the strong coupling regime.

\begin{figure}[ht]
	\centering
	\includegraphics[width=0.48\textwidth]{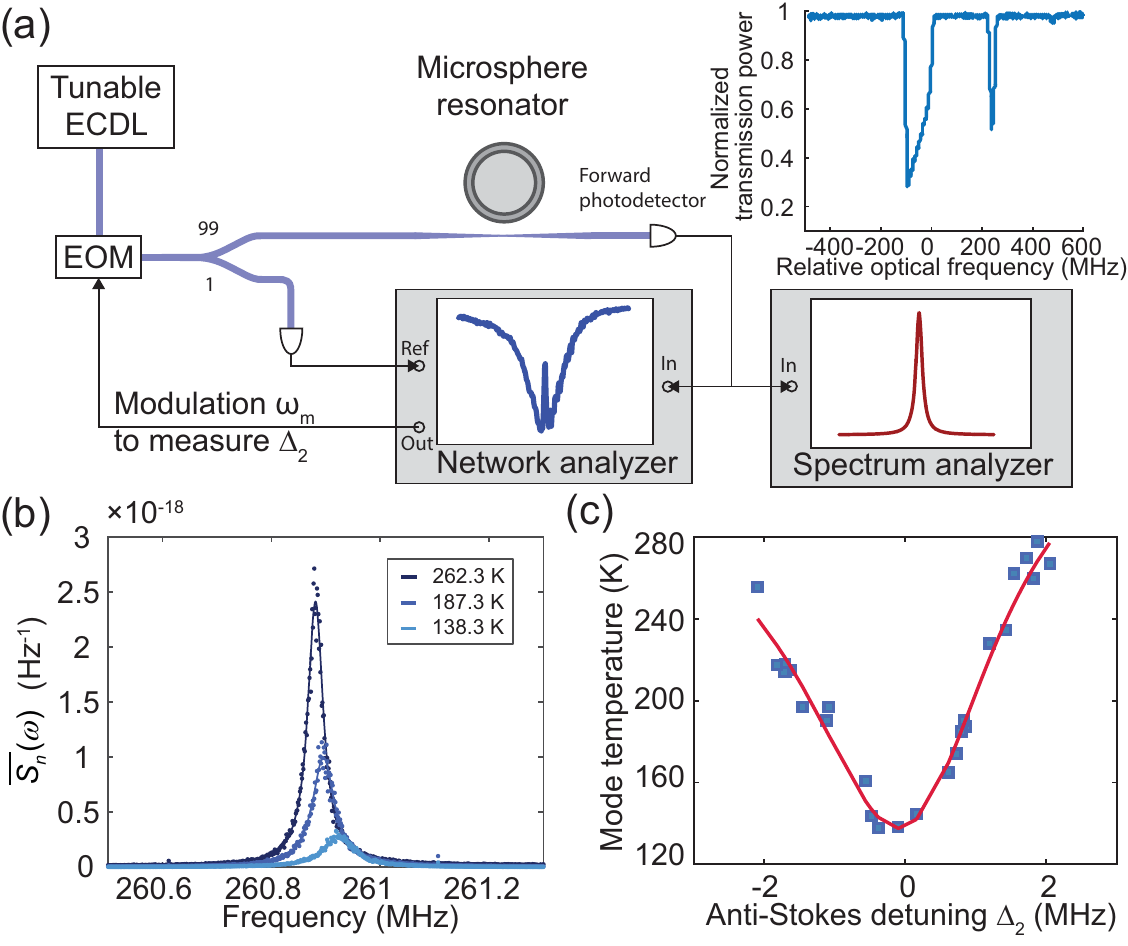}
	\caption{ 
		(a) Experimental setup used for investigating role of optical DoS in Brillouin cooling. A fiber-coupled tunable external cavity diode laser (ECDL) generates the pump field which is tuned to the ($\omega_1, k_1$) optical mode. The forward scattered signals are measured using a photodetector, electronic spectrum analyzer, and electronic network analyzer, using which the spectrum of the phonon mode and the anti-Stokes detuning $\Delta_2$ can thus be quantified. Top-right inset shows a fiber-transmission measurement of the two optical modes ($\omega_1, k_1$) and ($\omega_2, k_2$) during a laser sweep.  
		(b) The normalized phonon spectrum $\overline{S_{n}}(\omega)$ exhibits a cooling trend with increasing the pump laser. Solid lines are Lorentzian fits to the data.
		(c) The mode temperature $T_{\text{eff}}$ is presented as a function of $\Delta_{2}$. It shows that the cooled mode temperature does not follow a symmetric Lorentzian-type curve, which is indicative that the pump optical density of states also affects the cooling. 
	}
	\label{fig:2}
\end{figure}

Our experiments are performed in a whispering gallery resonator of optical Q~$>10^8$ with a diameter of about 150~$\mu m$ as shown in schematics Fig.~\ref{fig:1}(b) and Fig.~\ref{fig:2}(a). This device hosts two optical modes ($\omega_1, k_1$) and ($\omega_2, k_2$) that are phase-matched by a traveling phonon mode ($\Omega_B, q_B$).
The configuration of the optical modes is shown in $\omega-k$ space in Fig.~\ref{fig:1}(a). The resonator modes in consideration have loss rates $\kappa_{1}/2\pi = 1.6$ MHz and $\kappa_{2}/2\pi = 1.5$ MHz, while the phonon mode is at $\Omega_{B}/2\pi = 260.9$ MHz with loss rate $\Gamma_{B}/2\pi = 31.3$ kHz.
Thus, our system is in the resolved sideband regime ($\Omega_B \gg \kappa_1, \kappa_2 $).
An external cavity diode laser (ECDL) tunable over the C-band is used to pump the system. In this experiment we pump near 1555 nm with 1.1 mW launched in the waveguide. Light is coupled to the resonator via tapered optical fiber with external coupling rate $\kappa_{\text{ex}}$ to the resonator optical modes (Fig.~\ref{fig:1}(b)).
The pump is tuned to the ($\omega_1, k_1$) mode while spontaneous anti-Stokes scattering occurs to the ($\omega_2, k_2$) mode.
As shown in Fig.~\ref{fig:1}(a), Stokes scattering is simultaneously suppressed since no suitable optical mode is available. Forward scattered optical power from the resonator couples out via the tapered fiber and can be measured on a forward photodetector;
low-frequency transmission measurement shows the optical modes, while high-frequency measurement provides the frequency offset between pump and spontaneously scattered light. An electro-optic modulator (EOM) is used to generate a probe sideband, using which a network analyzer can monitor the detuning $\Delta_2 = (\omega_L + \Omega_B) - \omega_2$ between the scattered light and the ($\omega_2, k_2$) mode by means of Brillouin scattering induced transparency (BSIT) \cite{Kim2015a}. This induced transparency is illustrated in the network analyzer inset of Fig.~\ref{fig:2}(a).

The spectrum of the photocurrent $S_{II}(\omega)$ generated by pump and scattered light that appears on the forward detector is related to the symmetrized spectral density of the phonon mode $\overline{S_{bb}}(\omega)$ \cite{Chan2011,Clerk2010} via the relation
\begin{align}
	S_{II}(\omega)  = N + \left( \frac{4 \kappa_{\text{ex}}|g_0|^{2}\overbar{n}_{\text{cav}} }{\kappa_{2}^{2} + 4 \Delta_{2}^{2}} \right) \overline{S_{bb}}(\omega)       \label{eq:1} 
\end{align}
where the parameter $g_0$ is the optomechanical single-photon coupling strength, $\overbar{n}_{\text{cav}}$ is intracavity pump photon number in the ($\omega_1, k_1$) mode, and $N$ is the background noise floor.
Thus, the measured RF spectrum $S_{II}$ is a direct representation of the phonon mode spectrum $\overline{S_{bb}}(\omega)$ and is scaled according to the experimental parameters. To remove the dependence of this spectrum on the number of intracavity photons in the resonator $\overbar{n}_{\text{cav}}$ and the anti-Stokes field detuning $\Delta_2$, we perform a normalization to obtain   
$\overline{S_{n}}(\omega) = \frac{ (\kappa_{2}^2 + 4\Delta_{2}^2) }{4\kappa_{\text{ex}}|g_0|^2\overbar{n}_{\text{cav}}}S_{II}(\omega)$ (Fig.~\ref{fig:2}(b)).	
This $\overline{S_{n}}(\omega)$ is then proportional to $\overline{S_{bb}}(\omega)$  but with an additional background noise floor. 
We can then quantify the effective temperature $T_{\text{eff}}$ of the phonon mode by fitting its spectral linewidth $\Gamma_e$, and using the relation
$T_{\text{eff}} = \left( \Gamma_B/\Gamma_{e} \right) T_b$, where $T_b$ is the mechanical bath temperature (room temperature) and $\Gamma_B$ is the intrinsic linewidth of the phonon mode \cite{Schliesser2006}. We note that $\Gamma_e = \Gamma_B + \Gamma_{\text{opt}}$ where $\Gamma_{\text{opt}}$ is the optomechanical damping rate. 
The linewidth-temperature relation is derived from the equipartition theorem $k_{B}T_{\text{eff}} = \int_{-\infty}^{\infty}m_e \omega_{\text{o}}^2 \overline{S_{bb}}(\omega)d\omega$ \cite{MetzgerKarrai2004}. 

Since the pump and anti-Stokes optical fields have a well defined frequency offset $\Omega_B$, the detuning of the anti-Stokes field from the cavity resonance $\Delta_2$ can be simply modified by tuning the frequency of the pump.
Figure~\ref{fig:2}(c) shows the measured effective temperature of the phonon mode $T_{\text{eff}}$  as a function of $\Delta_{2}$. 
As we reduce $\Delta_2$ from 2 MHz to 0 MHz, the mode reaches a final effective temperature of $T_{\text{eff}}\sim 138.3$ K starting from 290~K. 
As anticipated, the effective temperature rises again to 255 K as we continue detuning to -2 MHz since the anti-Stokes optical density of states (DoS) modifies the scattering efficiency. 
The defining cooling curve in Fig.~\ref{fig:2}(c) does not follow a symmetric Lorentzian shape, since the pump detuning $\Delta_1$ is simultaneously modified in this experiment. This result shows a clear evidence of the dependence of the Brillouin cooling effect on the optical DoS of both the pump and the anti-Stokes optical modes. The measurable mechanical frequency $\Omega_B / 2\pi$ simultaneously tunes from 260.87 MHz to 260.96 MHz through the associated optical spring effect, as a function of detuning \cite{Hossein-Zadeh2007b}.  

\begin{figure}[ht]
	\centering
	\includegraphics[width=0.5\textwidth]{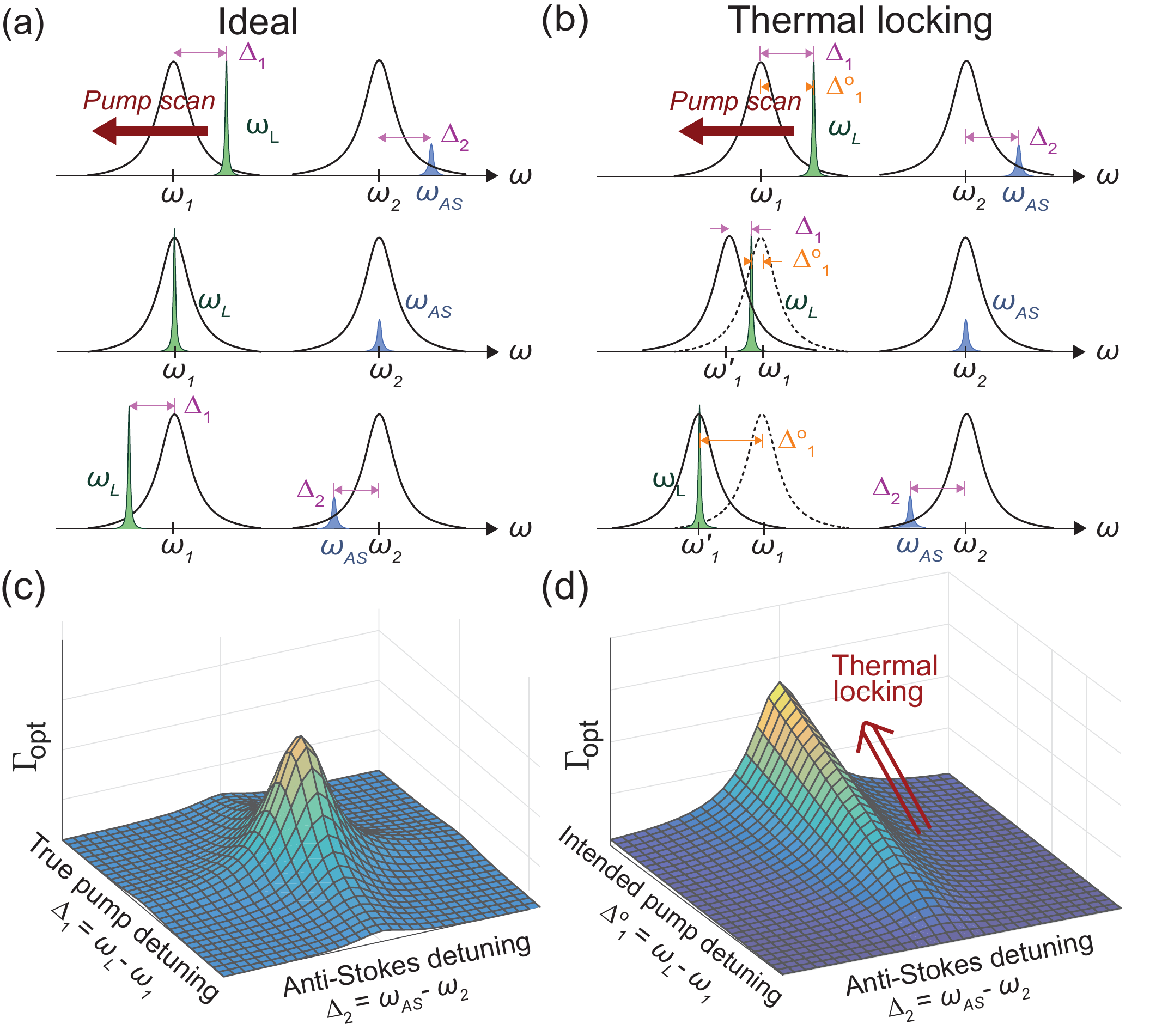}
	\caption{
		The role of two optical DoS on the optomechanical damping rate $\Gamma_{\text{opt}}$. 
		(a) Ideally, the pump (frequency $\omega_L$) and anti-Stokes (frequency $\omega_\text{AS}$) detuning track together due to the fixed acoustic mode frequency. 
		(b) Practically, thermal locking of the pump to its host optical mode causes the true anti-Stokes detuning to vary faster than the true pump detuning. $\Delta^\text{o}_1$ is the intended detuning while $\Delta_1$ is the true detuning.
		(c) Equation~\ref{eq:2} predicts that $\Gamma_{\text{opt}}$ is a Lorentzian function with respect to either of the true detunings $\Delta_1$ and $\Delta_2$. 
		(d) Due to thermal locking of the $\omega_1$ optical mode to the pump laser, the true pump detuning $\Delta_1$ does not track the intended pump detuning $\Delta^\text{o}_1$ (see (b)), resulting in a skewed cooling rate measured against $\Delta^\text{o}_1$. 
	} \label{fig:3}
\end{figure}

Let us now evaluate the characteristics of the phonon mode spectrum $\overline{S_{bb}}(\omega)$ under the influence of the cooling laser. The Hamiltonian formulation for such optomechanical systems with two optical modes has been presented previously \cite{Tomes2011,Agarwal2013}.
The resulting quantum Langevin equations enable derivation of the optomechanical damping rate $\Gamma_{\text{opt}}$ as follows:
	\begin{align}
		\Gamma_{\text{opt}} &=  \frac{4 \kappa_{2} |g_0|^{2} \overbar{n}_{\text{cav}}}{ \kappa_{2}^{2} + 4 \Delta_{2}^{2}}
		\textrm{~~where~~}
		\overbar{n}_{\text{cav}} = \frac{ 4 \kappa_{\text{ex}}  |\alpha_{\text{in}}|^{2} }{\kappa_{1}^{2} + 4 \Delta_{1} ^{2}}
		\label{eq:2}
	\end{align}
and where the pump detuning is $\Delta_{1} = \omega_{L} - \omega_{1}$.
Here $|\alpha_{\text{in}}|^2$ is the rate of pump photons arriving at the cavity from the waveguide.
Clearly, both the pump and anti-Stokes detunings, $\Delta_1$ and $\Delta_2$ affect the optical damping.
The optomechanical damping rate $\Gamma_{\text{opt}}$ given in Eq.~(\ref{eq:2}) leads us to envision a cooling scenario described in Fig.~\ref{fig:3}(c), where the cooling rate is dependent on detuning of the two optical signals to their respective modes $(\Delta_1,\ \Delta_2)$. A Lorentzian cooling response is thus expected on both independent detuning axes.
However, we know that the detuning parameters $\Delta_1$ and $\Delta_2$ are not independent, since the optical mode spacing and the phonon mode frequency are well determined. This is illustrated in Fig.~\ref{fig:3}(a) where the pump signal $\omega_L$ and anti-Stokes light $\omega_{AS}$ scan through the modes together since they must remain equidistant (the separation is defined by $\Omega_B$).

In reality, however, ultra-high-Q resonators do exhibit significant thermal tuning in response to pumping even in the sub-milliwatt regime.
This primarily takes the form of locking of the resonator mode to the pump laser \cite{Carmon2004}, and is illustrated in Fig.~\ref{fig:3}(b). The result is that the intended pump detuning $\Delta^\text{o}_1 = \omega_L - \omega_1$, i.e. the detuning from the original pump resonance ($\omega_1, k_1$), does not generate a Lorentzian shaped cooling rate but instead traces out a skewed response Fig.~\ref{fig:3}(d).
Experiments where measurable Brillouin cooling is to be observed require pump power in the $>10$ microwatt regime, and fall under this latter category.
As a result, the true pump detuning $\Delta_1 = \omega_L - \omega_1'$ for the locked mode at $\omega_1'$ is always lower than the intended detuning $\Delta^\text{o}_1 = \omega_L - \omega_1$.

\begin{figure}[ht]
	\centering
	\includegraphics[width=0.48\textwidth]{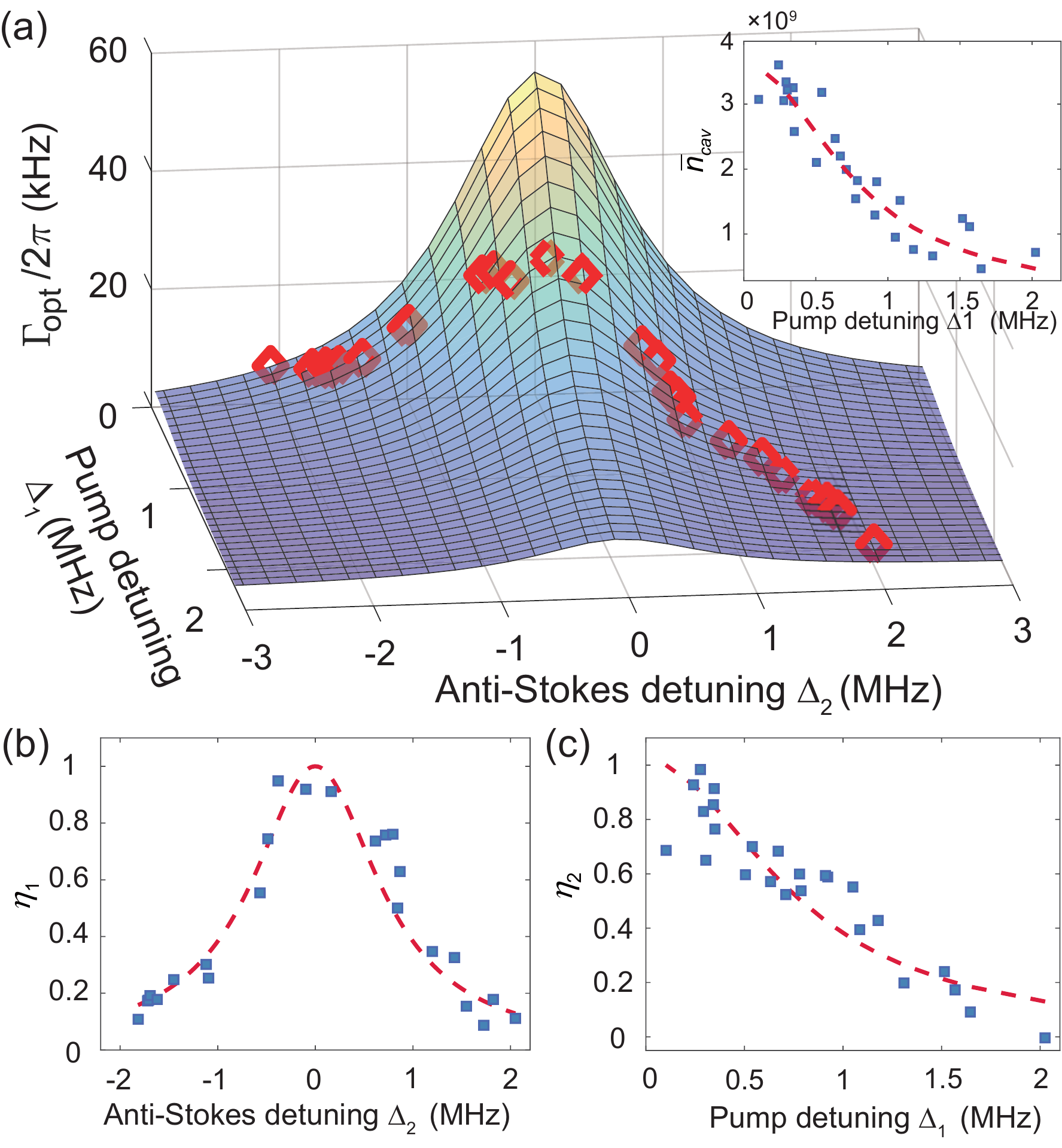}
	\caption{
		(a) Data points (red diamonds) are the measured optomechanical damping rate $\Gamma_{\text{opt}}$ a function of pump and anti-Stokes detuning $\Delta_{1}$ and $\Delta_{2}$. The 3d surface is the predicted result based on Eq.~(\ref{eq:2}), using the optomechanical single-photon coupling strength $g_0 /2\pi = 2.6 ~\text{Hz}$. 
		The inset shows the intracavity photon number $\overbar{n}_{\text{cav}}$ with respect to $\Delta_{1}$, as deduced from Eq.~(\ref{eq:2}).
		%
		(b) Dependence of $\eta_1$ and (c) $\eta_2$ (see text) on anti-Stokes detuning $\Delta_{2}$ and pump detuning $\Delta_{1}$, respectively. Red dashed lines are derived from Eq.~(\ref{eq:2}).
		}
		\label{fig:4}
\end{figure}

Figure \ref{fig:4}(a) shows the experimentally measured dependence of the optomechanical damping rate $\Gamma_{\text{opt}}$ on the true pump detuning $\Delta_{1}$ and the anti-Stokes detuning $\Delta_{2}$.
To relate the measurement to Eq.~(\ref{eq:2}), several experimental parameters are required.
The intracavity photon number $\overbar{n}_{\text{cav}}$ and the true pump detuning $\Delta_1$ can be inferred by measuring the power absorbed from the waveguide by the resonator. Specifically, $\Delta_1 =\sqrt{\frac{\kappa_1 \kappa_{\text{ex}}}{4} \big( \frac{P_{\text{wg}}}{P_{\text{in}}} - \frac{\kappa_1}{\kappa_{\text{ex}}} \big) }$ where $P_{\text{wg}}$ is the optical power in the waveguide and $P_{\text{in}}$ is the optical power absorbed by the resonator.
As mentioned above, the detuning $\Delta_2$ can be directly measured using BSIT \cite{Kim2015a}.
We can then predict the cooling rate $\Gamma_{\text{opt}}$ based on Eq.~(\ref{eq:2}) with all measurable parameters being known, except $g_0 /2\pi = 2.6 ~\text{Hz}$ which is extracted by fitting to the BSIT model. This prediction leads to the 3d surface shown in Fig.~\ref{fig:4}(a) and is in very good agreement with the experiment.
Since both pump and anti-Stokes detuning change together (see Figs.~\ref{fig:3}(a),(b)), the results are not dependent only on one optical DoS, but are affected by both simultaneously. As predicted, the true pump detuning varies far less than the anti-Stokes detuning due to thermal locking of the pump optical mode. 

\indent
We can now attempt to independently characterize the role of the optical DoS of either optical mode.
This can be achieved by describing normalized cooling rate $\zeta_1= {\Gamma_{opt}}/{\overbar{n}_{\text{cav}}}$, which is independent of the intracavity photon number $\overbar{n}_{\text{cav}}$ and $\Delta_1$, and normalized cooling rate $\zeta_2 = \Gamma_{opt}(\kappa_{2}^{2} + 4\Delta_{2}^{2})$ which is independent of the anti-Stokes detuning $\Delta_2$.
In addition, we define the dimensionless numbers $\eta_1 \triangleq \zeta_1 / \zeta_{1.\text{max}}$ and $\eta_2 \triangleq \zeta_2 / \zeta_{2.\text{max}}$ where $\zeta_{1.\text{max}}$ and $\zeta_{2.\text{max}}$ are the maximum value of $\zeta_{1}$ and $\zeta_{2}$ at resonances, i.e. at $\Delta_2 = 0$ and $\Delta_1 = 0$ respectively.
In Fig.~\ref{fig:4}(b) we see that the normalized measurement $\eta_1$ exhibits clear Lorentzian dependence on the anti-Stokes detuning $\Delta_2$. This result is similar to what is expected from resolved-sideband optomechanical cooling \cite{Park2009}.
Additionally, since we know the dependence of the intracavity photon number $\overbar{n}_{\text{cav}}$ on pump detuning $\Delta_{1}$, we are able to plot (in Fig.~\ref{fig:4}(c)) the normalized measurement $\eta_2$ against the pump detuning $\Delta_1$ directly, without being affected by the anti-Stokes mode detuning. There is thus a clear contribution of the DoS in both optical modes on the resulting phonon dissipation rate.

\begin{figure}[ht]
	\centering
	\includegraphics[width=0.48\textwidth]{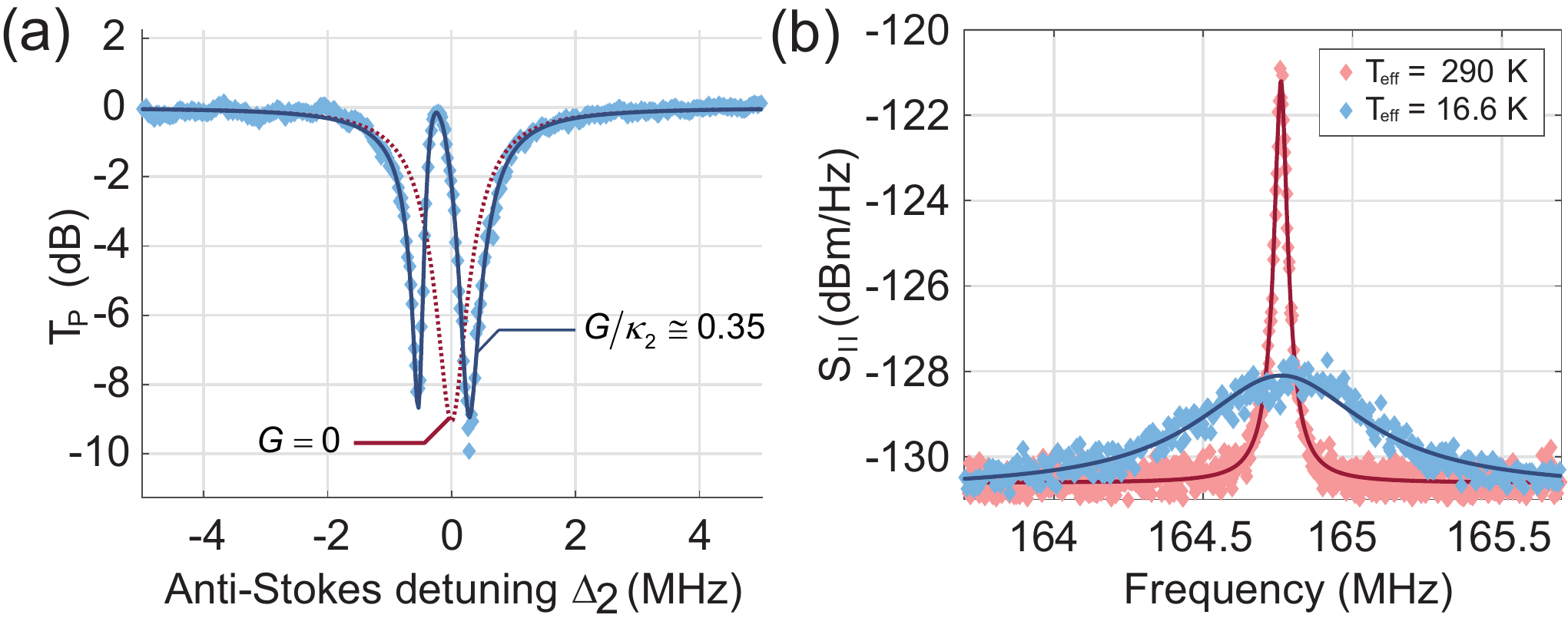}
	\caption{
		(a) Transmission of a probe laser tuned the anti-Stokes optical mode ($\omega_2, k_2$) is measured through the optical waveguide, while a fixed pump laser is tuned on the lower ($\omega_1, k_1$) optical mode. The red-dotted line indicates the dip induced by absorption into the anti-Stokes mode without optomechanical coupling ($G \sim 0$) i.e. zero pump.
		For non-zero optomechanical coupling, an EIT-like induced transparency (mode splitting) appears in the ($\omega_2, k_2$) mode due to destructive interference mediated by phonons \cite{Kim2015a}. The blue markers show experimental data while the blue-bold line shows the fit to theory.
		In this data, the ratio of the light-enhanced optomechanical coupling strength to the optical loss is $G / \kappa_2 \approx 0.35$, with $\kappa_2 /2\pi = 1.1$ MHz. 
		(b) The mechanical noise spectrum is measured at the photodetector $S_{II}$ for acoustic mode frequency $\Omega_B / 2\pi = 164.8$~MHz. The red data corresponds to phonon linewidth $\Gamma_B / 2\pi = 33.9$~kHz measured under very low optical pumping at room temperature. The blue data corresponds to $G / \kappa_2 \approx 0.35$, and shows $\Gamma_B / 2\pi = 592.3$~kHz (effective mode temperature of 16.6~K). }
	\label{fig:5}
\end{figure}

A special feature of two-mode optomechanical cooling systems is that the pump laser is resonant in the optical mode ($\omega_1, k_1$). These systems can thus support a very large amount of intracavity photons ($\overbar{n}_{\text{cav}} > 10^9$) compared to what is achievable in single-mode optomechanics \cite{{Park2009}, {Chan2011},  {SchliesserRiviereAnetsbergerEtAl2008}}.
As a result, the ratio of the light-enhanced optomechanical coupling strength $G = g_0 \sqrt{\overbar{n}_{\text{cav}}}$ to the optical loss rate $\kappa_2$ can be increased significantly, implying
that two-mode systems could potentially reach the strong coupling regime where $G \geq \kappa_2$, even at room temperature.
To explore this possibility, we arrange a system that has optical loss rates $\kappa_1 /2\pi \approx \kappa_2 /2\pi = 1.1$~MHz and acoustic loss rate $\Gamma_B / 2\pi = 33.9$~kHz with mechanical frequency $\Omega_B / 2\pi = 164.8$~MHz. 
Due to the optomechanical coupling, we anticipate the appearance of induced transparency in the anti-Stokes mode ($\omega_2, k_2$) \cite{Kim2015a}, which evolves into a mode splitting effect under strong coupling.
Figure~\ref{fig:5}(a) shows an experimental observation of this induced mode splitting that is generated due to the large circulating pump power ($\overbar{n}_{\text{cav}} = 10^{10}$) in the system. By fitting this observation to a mathematical model (Supplement of \cite{Kim2015a}) we can obtain $G / \kappa_2 \approx 0.35$, which is very close to the strong coupling regime.
The observed linewidth of the phonon mode is shown in Fig.~\ref{fig:5}(b), both before and after cooling. We see that the effective linewidth broadens to $\Gamma_e /2\pi = 592.3$ kHz which is comparable to the optical loss rate of the anti-Stokes mode $\kappa_2 /2 \pi = 1.1$ MHz. This result corresponds to a final mode temperature of 16.6 K. 
The cooling rate is expected to saturate at the optical loss rate $\kappa_2$ when the system reaches the strong coupling regime~\cite{DobrindtWilson-RaeKippenberg2008,RablKolkowitzKoppensEtAl2010}. Here, the measured effective mode temperature is very close to the estimated saturation temperature of 8.9~K.

In summary, we provided experimental verification that Brillouin cooling is dependent on two optical DoS, of both pump and anti-Stokes optical resonances. Additionally, we demonstrated that thermal locking of the pump laser causes the optomechanical damping rate $\Gamma_{\text{opt}}$ to be asymmetric with respect to the laser detuning. Finally, we show that resonant enhancement of the pump in a two-mode optomechanical cooling system enhances the cooling rate significantly. Our room temperature experimental results show that the strong coupling regime is well within reach, and may also pave the way to exploration of exploration of the saturation regime and beyond-saturation cooling techniques \cite{LiuXiaoLuanEtAl2013}.

\begin{acknowledgments}
	The authors would like to thank JunHwan Kim and Yin-Chung Chen for helpful technical discussions. This research was supported by the US Army Research Office (grant W911NF-15-1-0588) and the US Air Force Office for Scientific Research (grant FA9550-14-1-0217).
\end{acknowledgments}

\bibliography{biblio}

\end{document}